\documentstyle[prl,aps,multicol,epsf]{revtex}
\begin{document} 
\draft 
\title{Many-body band structure and Fermi surface of the Kondo lattice}

\author{R. Eder$^{1,2}$, O. Rogojanu$^2$, and G. A. Sawatzky$^2$}

\address{$^1$Institut f\"ur Theoretische Physik, Universit\"at W\"urzburg,
Am Hubland,  97074 W\"urzburg, Germany\\
$^2$Department of Solid State Physics, University of Groningen,
9747 AG Groningen, The Netherlands}
\date{\today}
\maketitle

\begin{abstract}
We present a theory for the single particle excitations and Fermi surface
of the Kondo lattice. Thereby we construct an effective Hamiltonian describing
the creation and propagation of single-particle-like charge fluctuations on 
an `RVB-background' of local singlets. The theory may be viewed as a 
Fermionic version of linear spin wave theory and is of comparable simplicity
so that the calculations for the strong-coupling limit can be performed 
analytically. We calculate the single particle spectral function for the 
`pure' Kondo lattice as well as for several extended versions: with a 
Coulomb repulsion between conduction and $f$-electrons, Coulomb repulsion 
between conduction electrons, and a `breathing' $f$-orbital. In all cases we 
study the evolution of the spectrum in going from the Kondo insulator to the 
Heavy electron metal. We compare our results to exact diagonalization of small
clusters and find remarkable agreement in nearly all cases studied. In the 
metallic 
case the $f$-electrons participate in the Fermi surface volume even when they 
are replaced by localized Kondo-spins and the number of bands, their dispersion
and spectral character, and the nontrivial (i.e. non-rigid band-like) doping 
dependence including a pronounced transfer of spectral weight are reproduced 
at least semiquantitatively by the theory.
\end{abstract} 
\begin{multicols}{2}

\section{Introduction}
The theoretical description of the Kondo lattice remains an outstanding 
problem of solid state physics. This model, or variations of it, may be 
viewed as the appropriate one for understanding such intensively investigated 
classes of materials as the heavy electron metals\cite{Stewart,Fulde,Bickers}, 
Kondo insulators\cite{Fisk} and possibly 
the recently discovered\cite{Huiberts} transition metal hydride-based
 switchable mirror compounds \cite{Ng,eps}.
The simplest model which incorporates the essential physics
may be written as
\begin{eqnarray}
H &=& 
\sum_{\bbox{k},\sigma} \epsilon_{\bbox{k}}
c_{\bbox{k},\sigma}^\dagger c_{\bbox{k},\sigma}^{} 
+ V \sum_{i,\sigma} (c_{i,\sigma}^\dagger f_{i,\sigma}^{} + H.c.)
\nonumber \\
&-& \epsilon_f \sum_{i,\sigma} f_{i,\sigma}^\dagger f_{i,\sigma}^{}
+ U_f\sum_{i} f_{i,\uparrow}^\dagger f_{i,\uparrow}^{}
 f_{i,\downarrow}^\dagger f_{i,\downarrow}^{}
\nonumber \\
&+&  U_{fc}\sum_{i,\sigma,\sigma'} f_{i,\sigma}^\dagger f_{i,\sigma}^{}
 c_{i,\sigma'}^\dagger c_{i,\sigma'}^{}
+ U_{c}\sum_{i} c_{i,\uparrow}^\dagger c_{i,\uparrow}^{}
 c_{i,\downarrow}^\dagger c_{i,\downarrow}^{}
\label{kondo1}
\end{eqnarray}
Here we consider the  minimal  model, where each
unit cell contains two orbitals, one of them for the
mobile conduction electrons
the other for the strongly correlated $f$-electrons. Then, 
$c_{i,\sigma}^\dagger$  ($f_{i,\sigma}^\dagger$)
creates a conduction electron ($f$-electron)
in cell $i$, 
and $\epsilon_{\bbox{k}}$$=$$\frac{1}{N}\sum_{i,j}
e^{i\bbox{k} \cdot(\bbox{R}_j - \bbox{R}_i)}\; t_{i,j}$
is the Fourier transform of the inter-cell
hopping integral $t_{i,j}$ for $c$ electrons.
For later reference we have already included a Coulomb 
repulsion $U_{fc}$ between
$f$ and conduction electrons in the same cell,
and a Coulomb repulsion $U_c$ between conduction electrons.
The latter two parameters are usually taken to be zero,
but as will be seen below our formalism
allows to take them into without any additional effort.
In the so-called symmetric case, $U_f=2\epsilon_f$
and the limit $U_f\rightarrow \infty$ (\ref{kondo1}),
can be reduced to its strong coupling limit\cite{SchriefferWolf}
\begin{equation}
H_{sc} =
\sum_{\bbox{k},\sigma} \epsilon_{\bbox{k}}
c_{\bbox{k},\sigma}^\dagger c_{\bbox{k},\sigma}^{} 
+ J \sum_i \bbox{S}_{i,c} \cdot \bbox{S}_{i,f}
\label{kondo2}
\end{equation}
where $\bbox{S}_{i,c}$ $(\bbox{S}_{i,f})$ denotes the
spin operator for conduction electrons ($f$-electrons)
in cell $i$ and $J=4V^2/U_f$.\\
While the impurity versions of these models, which retain only
a single $f$-site in a sea of conduction electrons, are well
understood\cite{Fulde,Bickers,Keiter,Coleman,Kuramoto,Hartmann} 
and are even amenable to exact 
solutions\cite{Andrei,Wilson}, 
much less is known about the lattice models.
One problem which by many is believed to be at the heart of the
solution is the way in which the more or less localized $f$-electrons,
which in the strong coupling theory are replaced by mere spin degrees 
of freedom, participate in the formation of the Fermi surface and the
 heavy  quasiparticle bands.
Experiments on heavy Fermion compounds\cite{Tailefer},
computer simulations of
Kondo lattices\cite{Moukuri,Tsutsui} and theoretical
considerations\cite{Tsvelik,Yamanaka} suggest that despite their
 frozen  charge degrees of freedom, the $f$-electrons participate in the
Fermi surface volume as if they were uncorrelated. In other words,
the experimental Fermi surface volume corresponds to the case
$U_f=U_{fc}=U_c=0$.
The limiting cases $V$$=$$0$ or $J$$=$$0$, which obviously
do not allow for participation of the $f$ electrons in the Fermi surface,
therefore represent singular points, so that a perturbation expansion
in the (small) parameters $V$ or $J$ may not be expected to give
meaningful results. Rather, the interaction between $f$-spins and conduction
electrons must be incorporated in a non-perturbative
way, in a similar manner as the single-impurity Kondo effect\cite{Wilson}.
It is the purpose of the present manuscript to present
a minimum effort theory for the Kondo lattice which is based
on this basic requirement and
shows how the nominal participation of the localized electrons in the
Fermi surface can be
understood even in the complete absence of any true hybridization.
We describe the system by an effective Hamiltonian for the
Fermion-like charge fluctuations  on top of  a strong coupling ground state,
and show that this treatment leads to remarkable agreement
with numerical results at least on energy scales which are
relevant to high-energy spectroscopy. 
We would also like to point out that the method of
calculation is similar in spirit
to the cell-perturbation method developed by 
Jefferson and coworkers\cite{Jefferson}.
A preliminary report has been published elsewhere\cite{eos}.\\
\section{Construction of the effective Hamiltonian}
\begin{figure}
\epsfxsize=12.1cm
\vspace{-9.5cm}
\hspace{-1.5cm}\epsffile{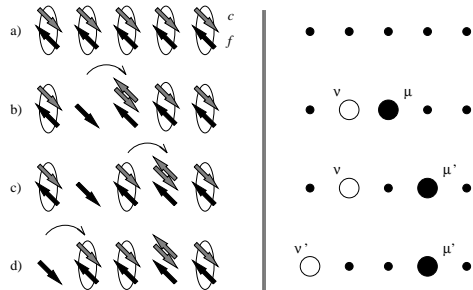}
\vspace{-5.0cm}
\narrowtext
\caption[]{Charge fluctuations and their propagation
(left panel) and their representation in terms of  model Fermions 
(right panel).}
\label{fig1} 
\end{figure}

\begin{figure}
\epsfxsize=8cm
\vspace{-0.0cm}
\hspace{0.5cm}\epsffile{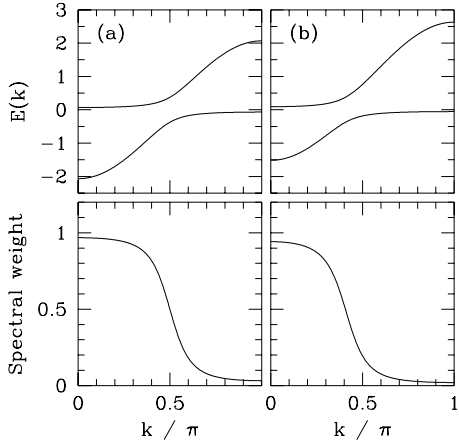}
\vspace{-2.0cm}
\narrowtext
\caption[]{Quasiparticle dispersion (top panels)
and dispersion of the $c$-like spectral weight 
along the lower band (bottom panels) for the 1D
strong coupling model with $J=0.5$,
$\epsilon_{\bbox{k}}= -2\cos(k)$. The density of
conduction electrons is $1$ in (a) and
$0.8$ in (b).}
\label{fig2} 
\end{figure}

As our starting point we choose the case of vanishing
inter-cell hopping $t_{i,j}$ (i.e. $\epsilon_{\bbox{k}}=0$).
The lattice problem then
reduces to single cell problems so that
we first discuss the eigenstates of a single cell
with $1$, $2$ and $3$ electrons.
The two 
electron ground state is a singlet with wave function
\begin{equation}
|\Psi_0^{(2)}\rangle = \;[\;
 \alpha f_{\uparrow}^\dagger f_{\downarrow}^\dagger
+ \frac{\beta}{\sqrt{2}}(c_{\uparrow}^\dagger f_{\downarrow}^\dagger
+ f_{\uparrow}^\dagger c_{\downarrow}^\dagger)
+\gamma c_{\uparrow}^\dagger c_{\downarrow}^\dagger 
\;]\;|vac\rangle
\label{two0}
\end{equation}
(we have suppressed the site index on the Fermion creation
operators).
The ground state wave function and energy
is then obtained by diagonalizing the $3\times 3$ matrix
\begin{equation}
H_2 = \left(
\begin{array}{c c c c c}
-2\epsilon_f+U_f&,&\sqrt{2}V&,&0\\
\sqrt{2}V&,& -\epsilon_f + U_{fc}&,&\sqrt{2}V\\
0&,&\sqrt{2}V&,&U_c
\end{array} \right).
\label{egs2}
\end{equation}
For the strong coupling model (\ref{kondo2}) the problem
becomes trivial with $\alpha$$=$$\gamma$$=$$0$, $\beta$$=$$1$,
the energy of the two-electron ground state is $-\frac{3J}{4}$.\\
The single or three electron 
states can be written as
\begin{eqnarray}
|\Psi_{\nu,\sigma}^{(1)}\rangle
&=& (\; \beta'_\nu f_{\sigma}^\dagger
+ \gamma'_\nu c_{\sigma}^\dagger\;)\;  |vac\rangle,
\nonumber \\
|\Psi_{\mu,\sigma}^{(3)}\rangle &=& \;(\;
\alpha''_\mu
c_{\sigma}^\dagger f_{\uparrow}^\dagger
f_{\downarrow}^\dagger
+ \beta''_\mu 
c_{\uparrow}^\dagger c_{\downarrow}^\dagger
f_{\sigma}^\dagger\;)\; |vac\rangle ,
\end{eqnarray}
and the wave functions and energies are obtained by
diagonalizing the matrices
\begin{equation}
H_1 = \left(
\begin{array}{c c c}
-\epsilon_f&,&V\\
V&,&0
\end{array} \right)
\end{equation}
\label{h1e}
and
\begin{equation}
H_3 = \left(
\begin{array}{c c c}
-2\epsilon_f+U_f+2U_{fc}&,&-V\\
-V&,&-\epsilon_f+U_c+2U_{fc}
\end{array} \right).
\label{h3e}
\end{equation}
For the strong coupling limit the index $\nu$ takes only one value
and we have $\gamma $$=$$\alpha  $$=0$,
and $\beta $$=$$\beta  $$=1$. Both the single and
three electron states have zero energy in this case.\\
For later reference, we also define the photoemission (PES)
and inverse photoemission (IPES) matrix elements
(here $\alpha$$=$$c,f$)
\begin{eqnarray}
r_{\alpha,\nu,\sigma} &=& 
\langle \Psi_{\nu,\bar{\sigma}}^{(1)}| \alpha_{\sigma}|
\Psi_{0}^{(2)}\rangle,
\nonumber \\
s_{\alpha,\mu,\sigma} &=& 
\langle \Psi_{\mu,\sigma}^{(3)}| \alpha_{\sigma}^\dagger|
\Psi_{0}^{(2)}\rangle.
\label{mats0}
\end{eqnarray}
They can be expressed in
terms of the wave functions defined above, e.g.
\begin{eqnarray}
\langle \Psi_{\nu,\bar{\sigma}}^{(1)}| c_{\sigma}|
\Psi_{0}^{(2)}\rangle &=& 
sign(\sigma)(\gamma \gamma'_\nu  + \frac{\beta 
\beta'_\nu }{\sqrt{2}}),
\nonumber \\
\langle \Psi_{\mu,\sigma}^{(3)}| c_{\sigma}^\dagger|
\Psi_{0}^{(2)}\rangle &=& \alpha \alpha''_\mu   - \frac{\beta 
\beta''_\mu  }{\sqrt{2}}
\end{eqnarray}
where the additional sign in the first equation
is due to our convention for ordering the two spin directions
in (\ref{two0})\cite{remark}.\\
We now return to the lattice problem and
consider the case of half-filling (i.e.
two electrons/unit cell, corresponding to the  Kondo insulator).
For vanishing $t_{i,j}$ the lattice ground state is simply the product
of $N$ single-cell ground states of the type (\ref{two0})
(see the state labeled (a) in Figure \ref{fig1}).
In the following, this state will be referred to
as  the vaccuum . Then, switching on $t_{i,j}$ produces  charge fluctuations 
in the vacuum state: an electron can 
\noindent 
hop from cell
$i$ to another cell $j$, leaving
the cell $i$ in a single electron eigenstate with number $\nu$,
and the cell $j$ in a three electron eigenstate with number $\mu$
(see state (b) in Figure \ref{fig1}).
In a further step an electron from the three-fold occupied
cell $j$ can hop to another neighbor $j $ leaving
cell $j$ in a two-hole eigenstate and $j $ in a three hole state
(see Figure \ref{fig1}c ) or, alternatively,
an electron can hop from
another neighbor $i $ into cell $i$, leaving $i$ in
a two electron state, $i $ in a single electron state.
Finally, an electron from $j$ can hop into $i$, leaving both
$i$ and $j$ in two-electron states. In this picture the inter-cell kinetic
energy may be viewed of as a perturbation which has a two-fold effect:
the pair creation of charge fluctuations and the propagation of these.
It therefore plays a completely analogous role as the
transverse part of the Heisenberg exchange in the linear spin wave theory
for the Heisenberg antiferromagnet, and in the following we want to
exploit this analogy. To that end we make the additional assumption
that a cell containing $2$ electrons must always be
in the $2$-electron ground state $|\Psi_0^{(2)}\rangle$.
This means that after a charge fluctuation has passed through
a given cell, the cell must return to the two-electron
ground state. In other words, under this constraint the propagating
charge fluctuations do not  leave a trace  of excited cells -
their propagation becomes coherent. Ways to relax this
constraint and take into account incoherent processes will be 
discussed below.\\
As a last important step, we note that the  pair creation process 
in Figure \ref{fig1}
changes the energy by $E_\nu + E_\mu - 2E_0$ (which is nothing but
the  conductivity gap  of a single cell), and that the switching
of e.g. a hole-like charge fluctuation
from species $\nu$ to species $\nu'$ changes the
energy by $E_{\nu'}^{(1)}-E_\nu^{(1)}$.
To keep track
of these changes in energy, we interpret the difference
$\epsilon_\nu$$=$$E_\nu^{(1)} - E_0^{(2)}$ as 
the `energy of formation' of
the hole species $\nu$ (and analogously for electron-like fluctuations). \\
We now define our restricted set of basis states:
\begin{equation}
|n_1\nu_1, n_2 \nu_2,\dots n_N\nu_N\rangle
= \prod_{i=1}^{i=N} |\Psi_{\nu_i}^{(n_i)} \rangle
\label{basis}
\end{equation}
with the side condition that $n_i=2$ automatically implies
$\nu_i=0$. In the following, we will diagonalize
the Hamiltonian in the subspace of the states (\ref{basis}).
To that end, we represent the basis states (\ref{basis}) in terms of
`model Fermions'\cite{sigrist}: if a cell with number $i$ is in the two-hole 
ground state we say it is empty; if the cell is in the $\nu^{th}$ single 
electron state with $z$-spin $\sigma$
we model this by the presence of a hole-like
Fermion, created by $a_{i,\nu,\sigma}^\dagger$; and if
there are three electrons forming the $\mu^{th}$ single
cell state with $z$-spin $\sigma$ we say that the cell is occupied by
an electron-like model Fermion, created by $b_{i,\mu,\sigma}^\dagger$.
Then, solving $H$ in the
restricted basis (\ref{basis}) obviously is equivalent to
diagonalizing the Hamiltonian
$H_{eff}$$=$${\cal P} H {\cal P}$ where
\begin{eqnarray}
H &=& 
\sum_{i,\sigma} \sum_{\nu} \epsilon_\nu\;
a_{i,\nu,\sigma}^\dagger\; a_{i,\nu,\sigma}^{}
 +
\sum_{i,\sigma} \sum_{\mu} \epsilon_\mu\;
b_{i,\mu,\sigma}^\dagger \; b_{i,\mu,\sigma}^{}
\nonumber\\
&+&\;\;\sum_{i,j,\sigma} \sum_{\nu,\mu} \;(\;
t_{i,\nu,j,\mu} \;\;b_{j,\mu,\sigma}^\dagger \;
a_{i,\nu,\bar{\sigma}}^\dagger
+ H.c.\;)\;
\nonumber \\
&+& \sum_{i,j,\sigma} \sum_{\nu,\nu'} 
\tilde{t}_{i,\nu,j,\nu'}\;\; a_{j,\nu,\sigma}^\dagger 
\;a_{i,\nu',\sigma}^{}
\nonumber \\
&+& \sum_{i,j,\sigma} \sum_{\mu,\mu'}
\tilde{\tilde{t}}_{i,\mu,j,\mu'}\;\; b_{j,\mu,\sigma}^\dagger 
\;b_{i,\mu',\sigma}^{}
\label{eff}
\end{eqnarray}
with
\begin{eqnarray}
t_{i,\nu,j,\mu} &=& \;\;\; t_{i,j} \;\;
\;r_{c,\nu,\sigma} \;\;\;s_{c,\mu,\sigma},
\nonumber \\
\tilde{t}_{i,\nu,j,\nu } &=& -t_{i,j} \;\;
\;r_{c,\nu',\sigma}^* \;\;\; r_{c,\nu,\sigma},
\nonumber \\
\tilde{\tilde{t}}_{i,\mu,j,\mu } &=& \;\;\; t_{i,j} \;\;
\;s_{c,\mu',\sigma}^*\;\;\; s_{c,\mu,\sigma}.
\label{mats}
\end{eqnarray}
Here ${\cal P}$ projects onto the subspace
of states where no site is occupied by more than one
Fermion. This kinematic constraint reflects the
fact that the state of a given cell must be unique.
Due to the product nature of the basis states (\ref{basis}), 
the evaluation of
the matrix elements of the inter-cell kinetic energy (\ref{mats})
reduces to the calculation of matrix elements between
products of no more than two single cell states. 
The matrix elements on the
r.h.s. of (\ref{mats})
are therefore
simply products of the $c$-like photoemission and inverse photoemission
matrix elements for a single cell (\ref{mats0}); this corresponds to the
quite intuitive picture that the propagation of a $c$-electron
is equivalent to photoemisson in one cell, and inverse photoemission
in the neighboring one. The respective matrix elements give the 
renormalization of the inter-cell hopping due to intra-cell 
(i.e. local) correlation effects. Also,
as long as the interaction between electrons contains only
intra-cell terms
the entire strong correlation physics obviously is 
completely taken care of
by the calculation of the single-cell states, and only enters via
the single cell energies $\epsilon_\nu = E_{\nu}^{(1)} - E_0^{(2)}$ and
$\epsilon_\mu = E_{\mu}^{(3)} - E_0^{(2)}$. Thus, while we are
presently only using Coulomb repulsions as intra-cell interactions,
Hund's rule exchange or electron phonon coupling
could also be treated in the same way.\\
Having computed the matrix elements and excitation energies
the most obvious next step then is (in analogy to linear
spin wave theory) to relax the constraint
enforced by ${\cal P}$, whereupon the Hamiltonian (\ref{eff})
is readily solved by Bogoliubov transformation.
This gives us the energies and dispersion, and for
the full Kondo lattice Hamiltonian we obtain
$4$ bands (we have two hole-like and two electron-like
model Fermions), whereas for the strong coupling version we have only
two. For the latter case, (\ref{eff}) takes the form
\begin{eqnarray}
H_{eff} &=& \frac{1}{2}\sum_{\bbox{k},\sigma}
[\;(-\epsilon_{\bbox{k}}+ \frac{3J}{2}) a_{\bbox{k},\sigma}^\dagger
a_{\bbox{k},\sigma}^{} +
(\epsilon_{\bbox{k}}+\frac{3J}{2} )
b_{\bbox{k},\sigma}^\dagger b_{\bbox{k},\sigma}^{} \;]
\nonumber \\
&-& \frac{1}{2}\sum_{\bbox{k},\sigma}
sign(\sigma)\; \epsilon_{\bbox{k}} \;
(b_{\bbox{k},\sigma}^\dagger a_{-\bbox{k},\bar{\sigma}}^\dagger
 + H.c.).
\label{stcham}
\end{eqnarray}
This is readily solved by the ansatz
\begin{eqnarray}
\gamma_{\bbox{k},1,\sigma} &=& \;\;
u_{\bbox{k},\sigma} b_{\bbox{k},\sigma}^{}
+ v_{\bbox{k},\sigma} a_{-\bbox{k},\bar{\sigma}}^\dagger
\nonumber \\
\gamma_{\bbox{k},2,\sigma} &=& -v_{\bbox{k},\sigma} 
b_{\bbox{k},\sigma}^{}
+ u_{\bbox{k},\sigma} a_{-\bbox{k},\bar{\sigma}}^\dagger
\label{ansatz}
\end{eqnarray}
and, introducing $\Delta$$=$$3J/2$, we
obtain the quasiparticle dispersion
\begin{equation}
E_\pm(\bbox{k}) = (1/2)\;[\; \epsilon_{\bbox{k}} \pm
\sqrt{ \epsilon_{\bbox{k}}^2+ \Delta^2  }\;],
\label{scdisp}
\end{equation}
shown in Figure \ref{fig2}a. 
At half-filling, particle-hole symmetry requires
the chemical potential to be zero, so that the
lower of the two bands (\ref{scdisp}) is completely filled, the
upper one completely empty. We note that
formally (\ref{scdisp}) is completely
equivalent to the hybridization
of a dispersionless  effective  $f$-level in the band center
with a free
\noindent 
electron band with dispersion $\epsilon_{\bbox{k}}$,
the strength of the nominal mixing element being $\Delta$.
This results in the  heavy , i.e. almost dispersionless
bands immediately above and below the Fermi energy in
Figure \ref{fig2} - the
slope of these bands decreases strongly with decreasing $J$.
It should be noted, however, that the 
resulting energy gap of $\Delta$ does {\em not} arise from
the formation of a bonding and antibonding combination of
$c$-like and $f$-like Bloch states, as in the hybridization model;
rather, this gap originates from the energy cost to break
two intra-cell singlets in the first step
of a charge fluctuation. 
This gap therefore is of a very similar nature as the energy gap in a 
superconductor: the minimum energy for moving an electron from
some site $i$ to a distant site $j$ (which by definition is the
single particle gap of the system) is two times the
energy required to break a pair (which may be either
a Kondo singlet or a Cooper pair: the first pair is broken
at site $i$, because one member of the pair is removed - the
second pair is broken at site $j$ because the surpuls electron
interferes with the pair formation around this site.
\section{Single particle Green s function}
To compute the full single particle spectral function
we need to resolve the  ordinary  electron creation
and annihilation operators
in terms of the model Fermions, $a$ and $b$. Taking into account
our basic assumption, namely that a single cell with $2$
electrons can only be in its ground state, we can expand
the electron annihilation operator
(where $\alpha=c,f$)
\begin{eqnarray}
\alpha_{i,\sigma} &=& \sum_\nu u_\nu 
|\Psi_{\nu,\bar{\sigma}}^{(1)} \rangle
\langle \Psi_{0}^{(2)}|
+ \sum_\mu v_\mu
|\Psi_{0}^{(2)}\rangle
\langle \Psi_{\mu,\sigma}^{(3)}|,
\nonumber \\
\alpha_{i,\sigma}^\dagger &=& \sum_\nu v_\mu^* 
|\Psi_{\mu,\sigma}^{(3)} \rangle
\langle \Psi_{0}^{(2)}|
+ \sum_\nu u_\nu^*
|\Psi_{0}^{(2)}\rangle
\langle \Psi_{\nu,\bar{\sigma}}^{(1)}|
\label{exp}
\end{eqnarray}
Taking matrix elements of both sides we readily
find that $u_\nu$$=$$r_{\alpha,\nu,\sigma}$,
$v_\mu$$=$$s_{\alpha,\mu,\sigma}^*$.
Next, we can replace
e.g. $|\Psi_{\nu,\bar{\sigma}}^{(1)} \rangle
\langle \Psi_{0}^{(2)}|
\rightarrow a_{i,\nu,\bar{\sigma}}^\dagger$ and thus have
the desired resolution of the photoemission operator.
Specializing to the strong coupling limit, the
annihilation operator for $c$-electrons takes the form
\begin{equation}
c_{\bbox{k},\sigma} = \frac{1}{\sqrt{2}}
(sign(\sigma) a_{-\bbox{k},\bar{\sigma}}^\dagger 
- b_{\bbox{k},\sigma}^{}).
\label{scphoto}
\end{equation}
Using (\ref{ansatz}) we can now resolve the annihilation
operator in terms of the quasiparticles, obtain its
matrix element and square it to
obtain the spectral weight in the lower
(i.e. occupied) band as
\begin{equation}
W = \frac{1}{2}(1 -
2 sign(\sigma) u_{\bbox{k},\sigma}v_{\bbox{k},\sigma}).
\end{equation}
Let us assume that $\epsilon_{\bbox{k}} \gg \Delta$ and
let $\bbox{k}$ be in the outer part of the Brillouin zone,
i.e. we assume that we are
 deep in the heavy portion  of the occupied band 
(see Figure \ref{fig2}a). Then, we find
\[
W = \frac{1}{4} \left( \Delta/\epsilon_{\bbox{k}}\right)^2 \ll 1
\]
i.e. the heavy band does have an extremely small spectral weight
(see the lower part of Figure \ref{fig2}a).
This will turn out to be of considerable importance in a minute.\\
To proceed to the
doped case, we need an expression for the electron number operator.
While at first sight this may appear a triviality,
we will now see
that one thereby runs into a rather deep-rooted problem, which
reflects the special features of the strong correlation problem.
In the  vacuum state  
the number of electrons (counting $c$ {\em and} $f$-electrons)
is $2N$ and the presence of an $a$-Fermion ($b$-Fermion)
decreases (increases) the electron number by $1$, so that for
the full Kondo lattice the
electron number operator should be simply
\begin{eqnarray}
N_e 
&=& 
\sum_{\bbox{k},\nu,\sigma}
a_{\bbox{k},\nu,\sigma}^{} a_{\bbox{k},\nu,\sigma}^\dagger
+ \sum_{\bbox{k},\mu,\sigma}
b_{\bbox{k},\mu,\sigma}^\dagger b_{\bbox{k},\mu,\sigma}^{} - 2N
\nonumber \\
&=& 
\sum_{\bbox{k},\sigma} \sum_{\mu=1}^4
\gamma_{\bbox{k},\mu,\sigma}^\dagger \gamma_{\bbox{k},\mu,\sigma}^{}
 - 2N.
\label{count}
\end{eqnarray}
For the strong coupling limit, we obtain in an analogous
fashion
\begin{eqnarray}
N_e =
\sum_{\bbox{k},\sigma} \sum_{\mu=1}^2
\gamma_{\bbox{k},\mu,\sigma}^\dagger \gamma_{\bbox{k},\mu,\sigma}^{}.
\label{count_sc}
\end{eqnarray}
The extra $-2N$ on the r.h.s. of (\ref{count}) simply cancels the
lowest of the $4$ bands obtained in the full Kondo lattice.
As will be seen below, this is a practically dispersionless
 lower Hubbard band  for the $f$-electrons, which is absent
in the strong coupling limit (or better: pushed to $-\infty$).
The Fermi surfaces of both models thus are completely
equivalent, and e.g. for the case of hole doping in $1D$ one would obtain a
Fermi momentum $k_F$$=$$\frac{\pi}{2}(\rho_c +1)$
(where $\rho_c<1$ denotes the density of conduction
electrons) in the lower hybridization band.
This implies that we have a Fermi surface which satisfies
a  nominal  Luttinger theorem, i.e. the $f$ -electrons are
treated as participating in the Fermi surface.
The physical origin, however, is the fact that
we have a density $1-\rho_c$ of  holes in the singlet background ,
which forms the vacuum for our treatment.
This results in a  hole pocket  centered on $k=\pi$ with Fermi momentum
$\frac{\pi}{2}(1-\rho_c)$. The resulting Fermi surface then is
{\em nominally} equivalent to a Luttinger Fermi surface {\em including} 
the $f$-electrons and obviously
this equivalence holds true irrespectively of
dimensionality and Fermi surface topology. We believe
that this is the reason why the $f$-electrons
{\em seem to} participate in the Fermi surface volume
despite the fact that they are localized. At half-filling, we do not
have a half-filled band of single-particle like mixtures
of $c$ and $f$-electrons - such a picture is obviously completely
wrong for the strong-coupling model. Rather, the half-filled ground state
should be viewed as an array of local singlets, where each
$f$-spin captures one conduction electron to form an 
immobile singlet. The system has a single particle gap because
removing an electron and re-inserting it at a far away site
breaks two of the local singlets, in a completely analogous fashion
as in a superconductor. Unlike a superconductor, the two
members of each pair do belong to different  species ,
whereby the number of one of the species, the $f$-electrons
is essentially independent of the total electron density.
Removing electrons from the system thus
produces  holes in the singlet background  and leaves behind
unpaired $f$-electrons. Unlike the actual $f$-electrons,
which are largely immobile (completely immobile for the
strong-coupling model) the  holes  are mobile,
because an $f$-electron can form a pair with a $c$-electron
from a nearby pair, thereby leaving another $f$-electron
unpaired.\\
However, there is still a major complication:
the electron number must also equal the $\bbox{k}$ and $\omega$
integrated photoemission weight, which is given by the expectation value
of the operator
\begin{equation}
N_e'  = 
\sum_{\bbox{k},\sigma} 
(c_{\bbox{k},\sigma}^\dagger c_{\bbox{k},\sigma}^{} +
f_{\bbox{k},\sigma}^\dagger f_{\bbox{k},\sigma}^{}).
\end{equation}
Inserting the expansion (\ref{exp}) of the $c$ and $f$ electrons
in terms of the model Fermions into this expression
it is easy to see that in general $N_e \neq N_e$. 
For example in the strong coupling
limit we find using (\ref{scphoto})
\begin{eqnarray}
N_e'&=& \frac{1}{2}
\sum_{\bbox{k},\sigma} 
[\; (a_{\bbox{k},\sigma}^{} a_{\bbox{k},\sigma}^\dagger +
b_{\bbox{k},\sigma}^\dagger b_{\bbox{k},\sigma}^{})
\nonumber \\
&\;&\;\;\;\;\; - sign(\sigma)
(b_{\bbox{k},\sigma}^\dagger a_{-\bbox{k},\bar{\sigma}}^\dagger 
+ a_{-\bbox{k},\bar{\sigma}}^{} b_{\bbox{k},\sigma}^{} )\;] + N.
\label{count1}
\end{eqnarray}
We thus arrive at the at first sight devastating
conclusion that counting the electrons in real
space on one hand and integrating the spectral weight
in $\bbox{k}$-space on the other hand give us different results
for the electron number. On the other hand,
this is not an indication for a qualitative
flaw in our theory, but has a very clear and simple
physical origin, namely the fact that
in a strongly correlated electron system spectral weight
and band structure are completely decoupled.
As an example, let us consider the band structure
shown in Figure \ref{fig2}a.
If one simply were to introduce a chemical potential
corresponding to the real-space electron count (\ref{count_sc})
the Fermi energy would necessarily cut into the  heavy  part
of the lower band.
Let us assume that the only effect of doping
were that $k_F$ progressively cuts deeper into the
heavy band portion of the otherwise completely rigid
quasiparticle band structure calculated above. Trivially, 
for the $1D$ chain the number of
occupied momenta, $N_{occ}$ would change with $k_F$ as
\[
\frac{ \partial N_{occ}}{\partial k_F} = 4.
\]
On the other hand, the integrated photoemission
weight, $N_{PES}$, would change as
\[
\frac{ \partial N_{PES}}{\partial k_F} = 4W = 
(\frac{\Delta}{\epsilon_{\bbox{k}_F}})^2.
\]
Introducing a simple chemical potential following (\ref{count_sc})
into an otherwise rigid quasiparticle band structure
thus inevitably leads to the breakdown of the sum rule for
the integrated photoemission weight. The breakdown of rigid-band behaviour
and the failure of a  simple electron count  are therefore
a completely natural consequence of the special feature
of strongly correlated electrons, namely that they may have quasiparticle 
weights which substantially deviate from unity. \\
In order to cope with this problem,
we try the simplest possible solution and 
enforce the consistency of  ordinary  
electron count and spectral weight integration
by adding both expressions, (\ref{count_sc}) and (\ref{count1}),
for the electron number to the Hamiltonian,
each one with a separate Lagrangian multiplier:
\begin{equation}
H \rightarrow H - \mu N_e - \lambda N_e'.
\end{equation}
The notion of  two chemical potentials  may seem
awkward at first sight, but as will be shown now, this approach 
results in a remarkable consistency with the numerical results.
Let us again consider the strong coupling limit. The
 spectral weight operator 
$N_e $ takes the same form as the kinetic
energy, but with the replacement $\epsilon_{\bbox{k}} \rightarrow 1$
or, equivalently, $t_{i,j} \rightarrow 1$.
The Hamiltonian thus becomes
\begin{eqnarray}
H_{eff} &=& \frac{1}{2}\sum_{\bbox{k},\sigma}
[-(\epsilon_{\bbox{k}}-\lambda) + \frac{3J}{2}+2\mu]\;
a_{\bbox{k},\sigma}^\dagger
a_{\bbox{k},\sigma}^{} 
\nonumber \\
&+&[(\epsilon_{\bbox{k}}-\lambda)+\frac{3J}{2} -2\mu ]\;
b_{\bbox{k},\sigma}^\dagger b_{\bbox{k},\sigma}^{}\; 
\nonumber \\
&-& \frac{1}{2}\sum_{\bbox{k},\sigma} sign(\sigma)
(\epsilon_{\bbox{k}}-\lambda) (b_{\bbox{k},\sigma}^\dagger
a_{\bbox{k},\bar{\sigma}}^\dagger + H.c. ),
\end{eqnarray}
so that, using again (\ref{ansatz}), we find the dispersion
\begin{equation}
E_\pm(\bbox{k}) = 
\frac{1}{2}[\;(\epsilon_{\bbox{k}} - \lambda) 
\pm \sqrt{ (\epsilon_{\bbox{k}} - \lambda)^2 +
\Delta^2}\;] - \mu.
\label{scdisp1}
\end{equation}
Using the representation of the spectral operators,
we obtain the momentum distribution/spin direction of the
conduction electrons:
\begin{equation}
n^c_{\bbox{k}} =
\frac{1}{2}( 1 -
\frac{\epsilon_{\bbox{k}} - \lambda}
{\sqrt{(\epsilon_{\bbox{k}} - \lambda)^2 + \Delta^2} }).
\label{nksc}
\end{equation}
Then,
for $|\epsilon_{\bbox{k}} - \lambda|\gg \Delta$ we may replace
the denominator on the r.h.s. by $-|\epsilon_{\bbox{k}} - \lambda|$
and obtain 
$n^c_{\bbox{k}}= \frac{1}{2}(1 - sign(\epsilon_{\bbox{k}} - \lambda))$.
This is simply the free-electron result.
For $\epsilon_{\bbox{k}} - \lambda=0$ on the other hand,
we find $n^c_{\bbox{k}}=1/2$.
Let us now discuss these results. To begin with,
the  real space chemical potential 
$\mu$ acts like a standard chemical potential,
which cuts into the  heavy  band and, as discussed above,
produces a Fermi surface consistent
with the nominal Luttinger theorem.
On the other hand, the chemical potential for the
spectral weight, $\lambda$, gives rise to a  pseudo Fermi surface 
for the conduction electrons, 
where $n^c_{\bbox{k}}$ drops sharply but continuously
from a value $\approx 1$ to nearly $0$ (see Figure \ref{fig2}b).
Since the integrated $c$-distribution must equal
the number of conduction electrons, it is clear that
$\lambda \approx \epsilon(k_F^0)$, with
$k_F^0$ the Fermi momentum for {\em unhybridized}
conduction electrons. Carrying on
the formal analogy of (\ref{scdisp1}) with a hybridization
gap picture, $\lambda$ obviously plays the role of an  on-site
energy  of the effective $f$ level, and therefore the line
$\epsilon_{\bbox{k}}=\lambda$ marks the
locus in $\bbox{k}$ space where the
strongly dispersive conduction-band  bends over  into the
nearly flat  heavy  band (see Figure \ref{fig2}b)
In other words, the band structure of the Kondo lattice
is equivalent to an effective $f$-level, which is pinned near
the  frozen core  Fermi energy for conduction electrons
of density $\rho_c$, and which mixes into the
conduction band with a matrix element
of strength $3J/4$. We note that this is
very much what one would expect intuitively:
in the limit $t\gg J$ the kinetic energy of the $c$-electrons
is by far the dominant energy contribution of the system.
This can be expressed as
\[
\langle H_{kin} \rangle = \int d\bbox{k}\;
 n^c_{\bbox{k}}\; \epsilon_{\bbox{k}}
\]
and obviously the minimum value compatible with the
Pauli principle is obtained by
the free electron distribution for the
conduction electrons. Then, in order to recover
a (small) additional energy $\sim J$ from the Kondo-hybridization,
the system will not sacrifice much kinetic energy, i.e.
$n^c(\bbox{k})$ will stay close to its free-electron shape.
\begin{figure}
\epsfxsize=8cm
\vspace{-0.5cm}
\hspace{0.5cm}\epsffile{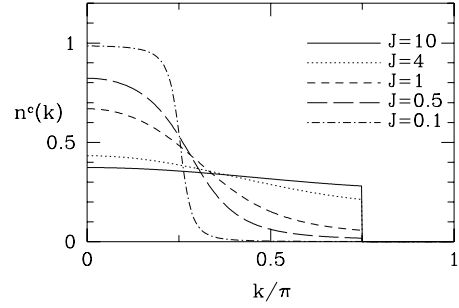}
\vspace{-3.5cm}
\narrowtext
\caption[]{Momentum distribution for the
strong coupling limit for different values of $J$, 
$\rho_c=0.5$.}
\label{fig3} 
\end{figure}
\noindent 
Figure \ref{fig3} shows  $n_c(\bbox{k})$ for a $1D$ chain
of the strong coupling model
with $\rho_c$$=$$0.5$,
evaluated numerically by solving self-consistently
for $\mu$ and $\lambda$, for different values of $J/t$.
For very large $J$ (which is an unphysical limiting case)
$n_c(\bbox{k})$ is nearly constant and drops to $0$ at
$k_F=3\pi/4$, the Fermi momentum for
hybridized conduction and $f$ electrons.
As $J$ is reduced, the
Fermi surface discontinuity shrinks more and more
(but stays finite at any $J$)
and the  pseudo Fermi surface  at $k_F^0$$=$$\pi/4$
starts to develop. This behaviour of $n_c(\bbox{k})$
is in almost quantitative agreement with the DMRG
results of Moukuri and Caron\cite{Moukuri}, which may
show the quality of our simple analytical calculation.\\
Having discussed the strong coupling version, we now turn to
the full Kondo lattice problem. Thereby we encounter
a new problem. In the strong coupling version, we have
introduced the  pseudo chemical potential  $\lambda$ in order
to enforce the consistency of real-space electron count
and integrated photoemission weight; clearly we will have to do the
same thing for the full Kondo lattice. For the full model,
however, there is an additional distinction to be made
because now we have $c$-like and $f$-like spectral weight,
and we need to make sure that the total spectral weight is
distributed between the two species in a proper version.
We will see that this enforces to introduce yet another
Lagrange multiplier which essentially governs the ratio
of $c$-like to $f$-like weight.\\
To that end, let us first construct an operator which counts the total 
number of $\alpha$-type electrons (where $\alpha=c,f$).
To begin with, we define the expectation values
of the electron numbers in the
single cell states:
\[
n_\alpha^{(i,\nu)} =
\langle \Psi_{\nu,\sigma}^{(i)}
| n_{\alpha,\uparrow} + n_{\alpha,\downarrow} |
\Psi_{\nu,\sigma}^{(i)} \rangle,
\] 
These expectation values are readily
computed from the single cell wave functions, e.g.
$n_f^{(2,0)}$$=$$2\alpha^2+\beta^2$.
By analogy with (\ref{count1}) we may thus write down the
following operator to count the total number of $\alpha$-electrons:
\begin{eqnarray}
N_{\alpha} &=& N n_{\alpha}^{(2,0)} +
\sum_{\bbox{k},\nu,\sigma} 
(n_{\alpha}^{(1,\nu)} - n_{\alpha}^{(2,0)})\;
a_{\bbox{k},\nu,\sigma}^\dagger a_{\bbox{k},\nu,\sigma}^{}
\nonumber \\
&+& \sum_{\bbox{k},\mu,\sigma} 
(n_{\alpha}^{(3,\mu)} - n_{\alpha}^{(2,0)})\;
b_{\bbox{k},\mu,\sigma}^\dagger b_{\bbox{k},\mu,\sigma}^{}
\label{count3}
\end{eqnarray}
Using  $n_f^{(i,\nu)} +n_c^{(i,\nu)}$$=$$i$ we find
$N_c$$+$$N_f$$=$$N_e$, with $N_e$ given by (\ref{count})
(as it has to be).
On the other hand by using the resolutions (\ref{exp})
of the $\alpha$ operators we can form the operators
$N_\alpha'$$=$$\sum_{\bbox{k},\sigma} \alpha_{\bbox{k},\sigma}^\dagger
\alpha_{\bbox{k},\sigma}$, which give the integrated
spectral weight for the $\alpha$-electrons alone.
We then replace the Hamiltonian
\begin{equation}
H \rightarrow
H - \mu N_e - \lambda (N_c' + N_f') - \lambda_f (N_f - N_f').
\label{hfinal}
\end{equation}
We determine $\lambda$ from the condition
$\langle N_c' + N_f'\rangle$$=$$N_e$ ($\lambda$ thus has precisely
the same meaning as for the strong coupling limit discussed above)
and $\lambda_f$ from the requirement
$\langle N_f - N_f' \rangle =0$. This implies that automatically
also  $\langle N_c - N_c' \rangle =0$, whence we
have reached  `species wise' consistency between 
real-space electron count and integrated
spectral weight.
\section{Comparison with exact diagonalization}
\begin{figure}[thb]
\epsfxsize=8.1cm
\vspace{-0.5cm}
\hspace{0cm}\epsffile{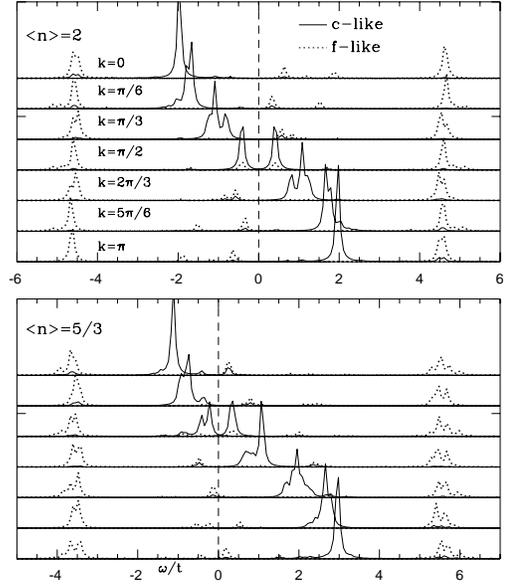}
\vspace{0cm}
\narrowtext
\caption{Single particle spectral function for the $1D$
Kondo lattice, obtained by Lanczos diagonalization of a $6$-unit-cell
system. Full lines (dashed lines) correspond to 
$c$-like ($f$-like) spectral weight. 
The vertical dashed line gives the Fermi energy $E_F$,
peaks to the right (left) of this line
correspond to electron creation (annihilation).
Parameter values are $U=8$, $\epsilon_f=4$, $V=1$,
$\epsilon(k)=-2\cos(k)$.}
\label{fig4} 
\end{figure}

\begin{figure}[htb]
\epsfxsize=8.1cm
\vspace{-0.5cm}
\hspace{-0.0cm}\epsffile{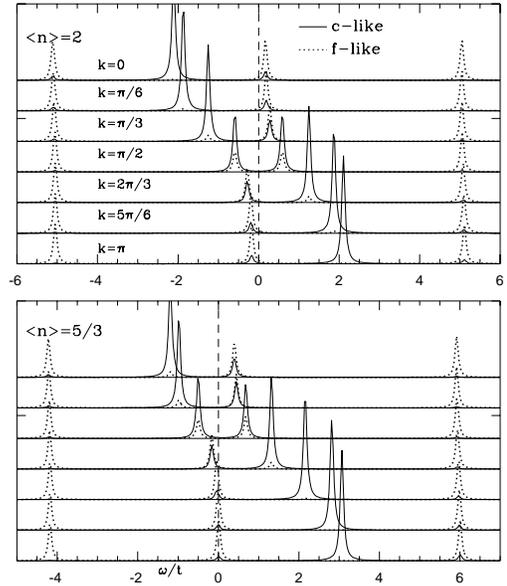}
\vspace{0cm}
\narrowtext
\caption{Theoretical spectrum for the same parameter values as
Figure \ref{fig4}.}
\label{fig5} 
\end{figure}

Using the eigenenergies obtained by diagonalizing the
effective Hamiltonian (\ref{hfinal}) with the self-consistently
determined values of the Lagrange multipliers $\lambda$
and $\lambda_f$ we obtain the band dispersions.
Combining the resolutions (\ref{exp})
of the $c$ and $f$-operators in terms of
the model Fermions and the eigenvectors obtained by
diagonalizing (\ref{hfinal}) we can compute
the photoemission and inverse photoemission matrix elements,
so that we can obtain the photoemission spectrum
\[
A_{\alpha}^{(-)}(\bbox{k},\omega)
= \frac{1}{\pi}
\Im \langle \Psi_0| \alpha_{\bbox{k},\sigma}^\dagger
\frac{1}{\omega + (H - E_0) - i0^+}
\alpha_{\bbox{k},\sigma}^{} | \Psi_0 \rangle
\]
and the inverse photoemisison spectrum
\[
A_{\alpha}^{(+)}(\bbox{k},\omega)
= \frac{1}{\pi}
\Im \langle \Psi_0| \alpha_{\bbox{k},\sigma}^{}
\frac{1}{\omega - (H - E_0) - i0^+}
\alpha_{\bbox{k},\sigma}^\dagger | \Psi_0 \rangle,
\]
where $\alpha=c,f$. Thereby
$|\Psi_0\rangle$ denotes the ground state wave function,
$E_0$ the ground state energy.
We have carried out this calculation for several $1$-dimensional
versions of the Kondo lattice and in the following
we compare them
to the results of Lanczos diagonalization. For the Lanczos 
studies we used a $6$-unit cell chain. To  simulate  longer chains,
we combine spectra calculated with periodic and
antiperiodic boundary conditions (see Ref. \cite{Tsutsui}
for a detailed discussion). While there is no 
rigorous justification for this procedure, inspection 
of the numerical spectra shows that one can obtain remarkably smooth
 band structures  in this way. \\
To begin with, Figure \ref{fig4} shows the 
Lanczos result for the single particle spectral 
function for a $1D$ chain of the  `pure'  Kondo lattice,
i.e. $U_{fc}$$=$$U_c$$=0$, Figure \ref{fig5}
shows the 
\noindent
result obtained from the theory
(similar Lanczos spectra with slightly different parameters
have also been obtained by Tsutsui {\em et al.}\cite{Tsutsui}).
To begin with,
in contrast to any band theory approach, our theoretical
spectrum correctly reproduces the $4$ well-distinguishable
bands  in the numerical spectra: the practically dispersionless
upper and lower Hubbard band , which have almost pure 
$f$-character, and the two hybridization bands, which 
\noindent
\begin{figure}[htb]
\epsfxsize=8.1cm
\vspace{-0.5cm}
\hspace{-0.0cm}\epsffile{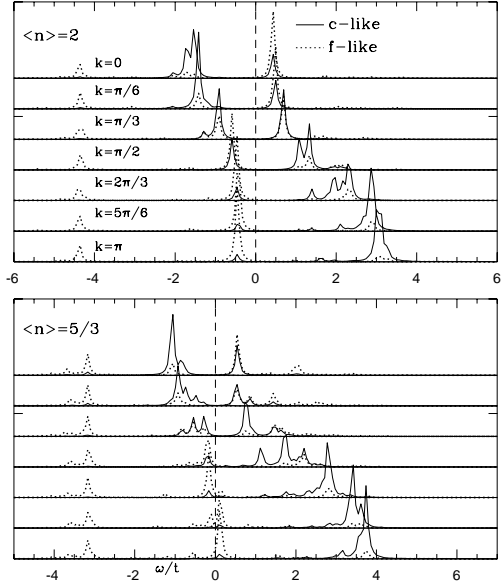}
\vspace{0cm}
\narrowtext
\caption{Lanczos spectrum for $U_f=4$, all other parameter values
as in Figure \ref{fig4}.}
\label{fig6} 
\end{figure}
\noindent
resemble the strong coupling-result (\ref{scdisp}). 
The basic idea of our approach, namely to broaden the
ionization and affinity states of a single cell into
bands thus obviously works quite well. 
The spectra nicely show the  Pseudo Fermi surface  for the
$c$-electrons: while the  true  Fermi momentum in
the doped case is at $k_F$$=$$5\pi/6$, where the heavy band
intersects $E_F$ (although this is not easily recognized in
the theoretical spectra), there is a pronounced 
transfer of $c$-like spectral weight from below to
above $E_F$ at $k_F^0=\pi/2$.
This means that doping shifts the drop
of the momentum distribution for the $c$-electrons 
(which is nothing but the integrated $c$-weight below $E_F$) 
from 
\begin{figure}[htb]
\epsfxsize=8.1cm
\vspace{-0.5cm}
\hspace{-0.0cm}\epsffile{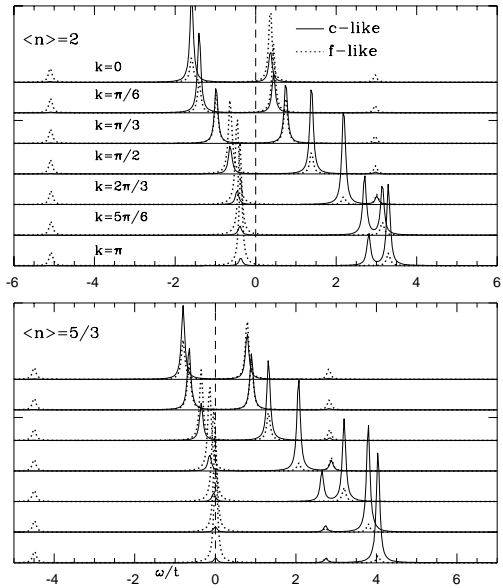}
\vspace{0cm}
\narrowtext
\caption{Theoretical spectrum for the parameter values
as in Figure \ref{fig6}.}
\label{fig7} 
\end{figure}
\noindent
$\pi/2$ to $\pi/3$ - this is 
preciseley the  pseudo Fermi surface  discussed above
for the strong coupling limit (see Figure \ref{fig2}).
The theoretical spectra somewhat overestimate
the weight of the  heavy  $f$-like portions in the hybridization
bands, moreover the position of the $f$-like Hubbard
bands is shifted to slightly too high (binding) energies.
On the other hand, the  Fermiology  is reproduced
quite well. It is interesting to note
that qualitatively identical results
for smaller values of $U_f$ and $\epsilon_f$ and lower electron filling
have been obtained by Tahvildar-Zadeh {\em at al.}\cite{tahvildar}
using the $D\rightarrow \infty$ technique - quite obviously there
is overall agreement between different numerical approaches,
for quite different parameters values and fillings
(and our analytical results).\\
Figure \ref{fig6} shows a more asymmetric case, more precisely
we set $\epsilon_f$$=$$U$. As expected, the upper Hubbard band for the
$f$-electrons is virtually
absent. Interestingly, the theoretical spectrum 
shows a very weak  band  at $\approx 3t$ above the Fermi 
energy, 
whose intensity shows a weak increase with doping. 
The Lanczos spectrum for half-filling
shows some diffuse $f$-weight roughly in this area,
and even some indication of  peaks  in the doped case.
Next, the dispersion of the hybridization bands
shows a quite pronounced asymmetry between photoemisison
and inverse photoemission, which is nicely reproduced
by the theory. The explanation is simple: due to the
reduced energy of the  pure $f$-state 
$f_{\uparrow}^\dagger f_{\downarrow}^\dagger|vac
\rangle$ the weight of this state in the
single cell ground state for two electrons
$|\Psi_0^{(2)}\rangle$  must increase. The weight
of the two remaining states therefore must decrease, so that the
overall probability to find a $c$-electron in the ground state 
is reduced
(this manifests itself also by the significantly
smaller integrated $c$-weight
in the photoemission spectrum).
Since only the $c$-electrons can hop between cells,
the band-width seen in photoemission therefore will be
reduced. Next, since the
$c$-weight which disappears from the photoemission part must
reappear in the inverse photoemission spectrum, it follows from
analogous considerations that the dispersion in the
IPES part will be enhanced. As was the case
in the symmetric case, doping causes a shift of the
chemical potential into the lower hybridization band,
but there is now also some change of the spectral character of the upper
hybridization band near $k=\pi$: there the $c$-character increases,
the $f$-character decreases with doping, and the theoretical
spectra obviously do correctly reproduce this trend.\\
Next, we study a more complicated form of the hopping
term for conduction electrons. More precisley, we introduce
a hopping integral between second nearest neighbors, and to have a
pronounced effect we choose this to be of equal magnitude, but opposite
sign as the nearest neighbor hopping. In other words,
the conduction electrons now have the dispersion relation
\begin{equation}
\epsilon_{k} = -2t( \cos(k) - \cos(2k)).
\label{newdisp}
\end{equation}
\begin{figure}[htb]
\epsfxsize=8.1cm
\vspace{-0.5cm}
\hspace{-0.0cm}\epsffile{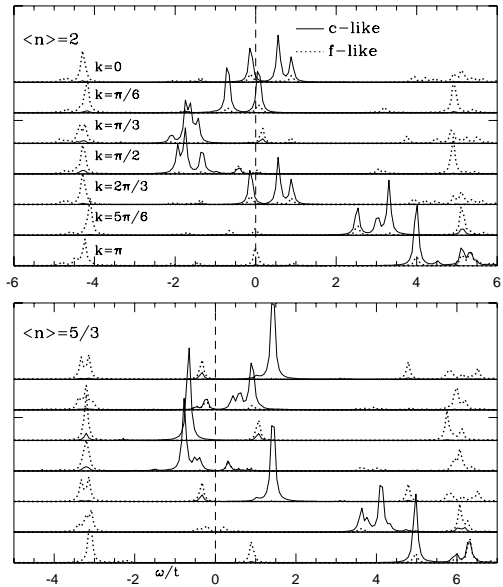}
\vspace{0cm}
\narrowtext
\caption{Lanczos spectrum for conduction
electrons with the dispersion (\ref{newdisp}), all other parameter 
values as in Figure \ref{fig4}.}
\label{fig8} 
\end{figure}
\noindent
For the band fillings under consideration, a system of noninteracting
electrons thus would have $4$ Fermi points rather
than $2$, and we want to see if our simple 
`rule of thumb' deduced above for the band structure
in the strong coupling limit continues to be valid also in
this more complicated situation. Namely the band structure of the Kondo 
lattice should be qualitatively given by the one obtained for mixing
with a dispersionless $f$-level, that is pinned to the
`frozen core' Fermi energy of the non-hybridizing conduction
band, see Figure \ref{fig2}.
Then, figure \ref{fig8} shows the Lanczos result for the single particle
spectral function and Figure \ref{fig9} shows the theoretical
\begin{figure}[htb]
\epsfxsize=8.1cm
\vspace{-0.5cm}
\hspace{-0.0cm}\epsffile{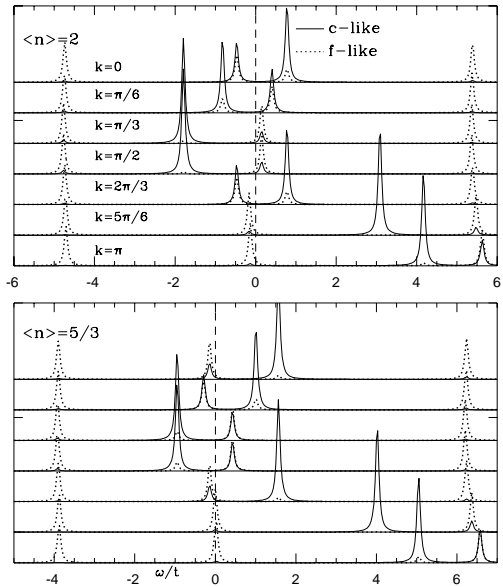}
\vspace{0cm}
\narrowtext
\caption{Theoretical spectrum for the parameter values
as in Figure \ref{fig8}.}
\label{fig9} 
\end{figure}
\noindent
data. It is quite obvious that the theory remains
valid also in this case. The relatively complicated
band structure predicted by the  rule of thumb  indeed can be seen quite
well in the Lanczos spectra: the $s$-shaped conduction band,
with a local maximum at $k=0$, which crosses from inverse
photoemission and back again and 
thereby mixes with the dispersionless, low intensity 
$f$-band (it should be noted that the low intensity $f$-like peaks at
$5\pi/6$ and $\pi$ in the Lanczos spectra for $\langle n\rangle=2$
are actually on the photoemission side). With doping away from
half-filling the heavy band near $k=\pi$ develops a
Fermi edge, and the {\em two } Pseudo Fermi surfaces for the
$c$-electrons at $\pi/6$ and $2\pi/3$
shift as expected for free, nonhybridizing $c$-electrons. 
In the doped case, the flat low intensity band which skims below $E_F$
and crosses $E_F$ at $k_F=5\pi/6$ can be identified particularly well. Also
the  heavy  part in the inverse photoemission at $k=\pi/3$ and $\pi/2$
which is predicted by the theory can be identified reasonably well.
Again, the theoretical spectra somewhat overestimate the
spectral weight of these 
heavy band portions, and the relative
position of the lower Hubbard 
band is not correct.
Our simple picture of the Fermiology, however, obviously remains valid also
with this more complicated Fermi surface topology.\\
We proceed to the case of non-vanishing $U_{fc}$, i.e.
a Coulomb repulsion between $f$ and conduction electrons.
The Lanczos spectra for this case are shown in Figure
\ref{fig10}, the theoretical spectra in Figure \ref{fig11}.
The main effects of
$U_{fc}$ is to open a wider gap at half-filling, to
enhance the weight of the  heavy bands  and to
shift the Hubbard-bands to higher energy. All in all,
the total width of the spectrum is now $\approx 12 = U_f+U_{fc}$
rather than $U_f$ as it used to be in the preceding cases.
As compared to
the preceding cases, the Hubbard bands moreover acquire an appreciable 
dispersion, and doping causes a pronounced spectral weight transfer
from the upper to the lower Hubbard band. 
The widening of the
gap at half-filling appears hard to understand at first sight, 
because $U_{fc}$ by itself 
does not increase the energy cost for a charge fluctuation: 
transferring a $c$-electron from one cell to another
(see the process Figure \ref{fig1}a $\rightarrow$ \ref{fig1}b)
does not change the energy due to $f$-$c$ repulsion as long
as the $f$-occupation is nearly constant at $1$ (which is almost
certainly the case for the large $\epsilon_f$ we are using).
It is straightforward to see, however,
that the opening of the gap is due to a loss of
{\em kinetic energy}: switching on $U_{fc}$
increases the energy of
the state $\frac{1}{\sqrt{2}}(c_{\uparrow}^\dagger 
f_{\downarrow}^\dagger + f_{\uparrow}^\dagger c_{\downarrow}^\dagger)$
by $U_{fc}$, so that 
the energy difference relative to the other two basis states
is decreased by this amount (see the Hamilton matrix
(\ref{egs2})).
This means that charge fluctuations
in the two-electron ground state are enhanced, whence the
kinetic energy for this state becomes more negative.
On the other hand,
this mechanism for enhancing the charge fluctuations is operative
neither for one nor for three electrons in a cell
(see the respective Hamilton matrices (\ref{h1e}) 
\begin{figure}[htb]
\epsfxsize=8.1cm
\vspace{-0.5cm}
\hspace{-0.0cm}\epsffile{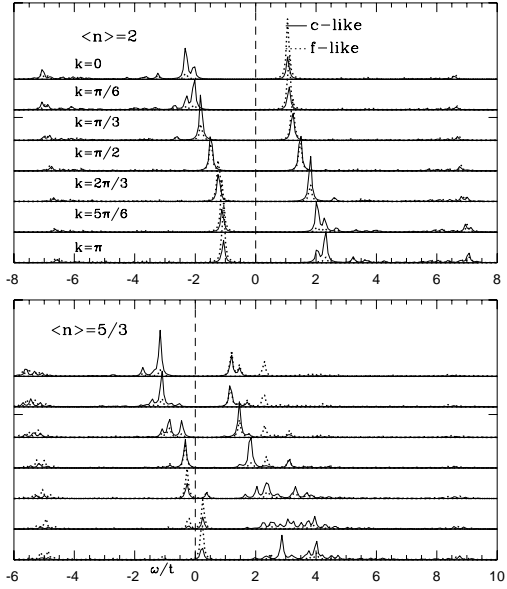}
\vspace{0cm}
\narrowtext
\caption{Lanczos spectrum with $U_{fc}$$=$$4$,
all other parameter values as in Figure \ref{fig4}.}
\label{fig10} 
\end{figure}
\noindent
and (\ref{h3e}),
so that there is no gain in kinetic energy in these cases.
The charge fluctuation thus
is accompanied by a net loss of kinetic energy, which in turn
results in a larger charge gap. 
Moreover, it follows from the expression (\ref{nksc}) for the
$c$-type spectral weight of the lower hybridization band
that the distance in $\bbox{k}$-space over which the
weight drops is approximately given by $\Delta_c/v_F$,
with $\Delta_c$ the charge gap and $v_F$ the Fermi velocity.
It is thus immediately obvious that an increase of the charge gap
(for whatever reason) will give a `more homogeneous' 
$c$-weight along the hybridization bands.
While at half-filling
the agreement between Lanczos and theoretical 
\begin{figure}[htb]
\epsfxsize=8.1cm
\vspace{-0.5cm}
\hspace{-0.0cm}\epsffile{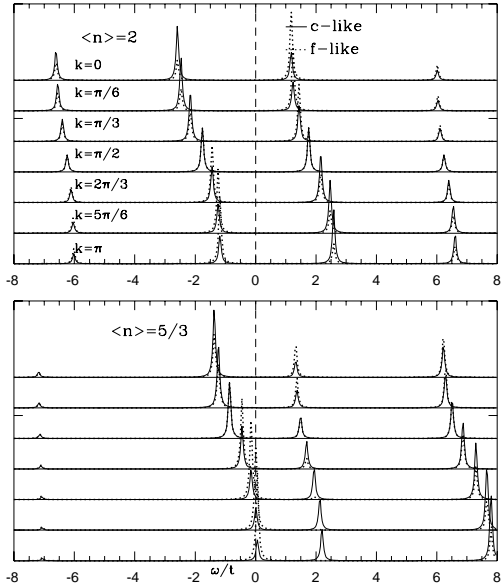}
\vspace{0cm}
\narrowtext
\caption{Theoretical spectrum for the parameter values
as in Figure \ref{fig10}.}
\label{fig11} 
\end{figure}
\noindent
\begin{figure}[htb]
\epsfxsize=8.1cm
\vspace{-0.5cm}
\hspace{-0.0cm}\epsffile{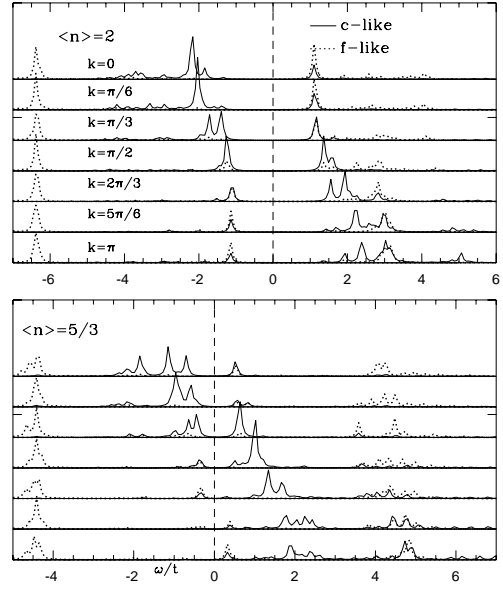}
\vspace{0cm}
\narrowtext
\caption{Lanczos spectrum with $U_{c}$$=$$4$,
all other parameter values as in Figure \ref{fig4}.}
\label{fig12} 
\end{figure}
\noindent
spectra is very good,
the situation changes for the doped case. The lower of the two
hybridization bands is still well described, but the
weights and dispersions of the other bands, as well
as the weight shift between the Hubbard bands are not
reproduced well. Actually the Lanczos spectra show a rather
strong reduction of the gap between the two central bands
upon doping - we believe that this indicates a rather profound
doping-induced reconstruction of the electronic structure
and since our theory corresponds 
more or less to an `expansion 
around the Kondo insulator' it may not be
expected to reproduce this.\\
\begin{figure}[htb]
\epsfxsize=8.1cm
\vspace{-0.5cm}
\hspace{-0.0cm}\epsffile{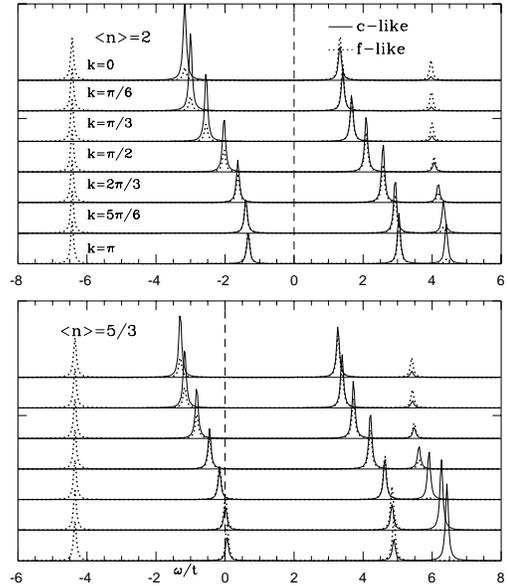}
\vspace{0cm}
\narrowtext
\caption{Theoretical spectra for the parameters of
Figure \ref{fig12},
all other parameter values as in Figure \ref{fig4}.}
\label{fig13} 
\end{figure}
\noindent
The situation is actually quite similar if we introduce
a Coulomb repulsion between conduction
electrons\cite{schork,khaliullin,itai}, $U_c$, see Figure \ref{fig12} and
\ref{fig13}. Again, this opens a wider quasiparticle gap in the
spectrum and the  hybridization bands  now have a
predominant $c$-character over their entire width.
At half-filling the upper hybridization band intersects 
and mixes with the
upper Hubbard band for the $f$-electrons.
The upper and lower Hubbard band for the $f$-electrons remain 
unaffected and are dispersionless, the lower one has practically pure
$f$-like character. At half-filling, there is again
good agreement between Lanczos and theory but for the
doped case the situation is again different.
The Lanczos spectra show a rather dramatic collapse of
the gap between the hybridization bands, the upper
hybridization band, which intersected the upper Hubbard band at 
half-filling is now at least $2t$ below. Again, this
suggests a strong doping induced
reconstruction of the entire electronic structure, and
our theory naturally fails to reproduce this.\\
The effect of $U_c$ at half-filling
can be understood already in the context
of the strong coupling model limit, where we simply would have to
replace $\Delta\rightarrow \Delta + U_c$.
The more homogeneous $c$-character of the
hybridization bands then follows in an analogous fashion
as for $U_{fc}$.\\
As a last example, we turn to a somewhat exotic version of the
Kondo lattice model, where the magnitude of the
$c$-$f$ hybridization depends on the occupation of the
$f$-level. More precisely, we replace
\[
V c_{i,\sigma}^\dagger f_{i,\sigma}^{} \rightarrow
V_1 c_{i,\sigma}^\dagger f_{i,\sigma}^{} 
f_{i,\bar{\sigma}}^\dagger f_{i,\bar{\sigma}}^{}
+ V_2 c_{i,\sigma}^\dagger f_{i,\sigma}^{} 
f_{i,\bar{\sigma}}^{} f_{i,\bar{\sigma}}^\dagger
\] 
with $V_1 \gg V_2$.
Such a model may be relevant\cite{eps} to describe the 
recently discovered\cite{Huiberts} transition metal
hydrides with  switchable mirror  properties, such as
YH$_3$. In this case, the Yttrium $4d$ electrons
would play the role of the  conduction electrons ,
the Hydrogen corresponds to the  $f$-electrons .
The conditional hopping term is supposed to
describe the relaxation of the orbital wave function
on Hydrogen as a function of electron
occupation\cite{eps}. This is manifested e.g. by the dramatically larger
radius of the free $H^-$ ion as compared to the neutral free
$H$ atom (for a detailed discussion see Ref. \cite{eps},
for a discussion of such an `orbital Kondo effect' in the
context of cuprate superconductors see Refs.\cite{Hirsch,Penc}). 
Here we are not so much interested in the details of
the correspondence with YH$_3$, but rather in the applicability
of our theory to this model. 
The conditional hopping is easily incorporated by
replacing the Hamiltonian (\ref{egs2}) 
for two electrons in a cell by
\begin{equation}
H_2  = \left(
\begin{array}{c c c c c}
-2\epsilon_f+U_f&,&\sqrt{2}V_1&,&0\\
\sqrt{2}V_1&,& -\epsilon_f + U_{fc}&,&\sqrt{2}V_2\\
0&,&\sqrt{2}V_2&,&U_c
\end{array} \right)
\end{equation}
and analogous replacements in the single and three-electron
subspaces. It is quite obvious, that
the conditional hopping will lead to
a dramatic increase of the 
\begin{figure}[htb]
\epsfxsize=8.1cm
\vspace{-0.5cm}
\hspace{-0.0cm}\epsffile{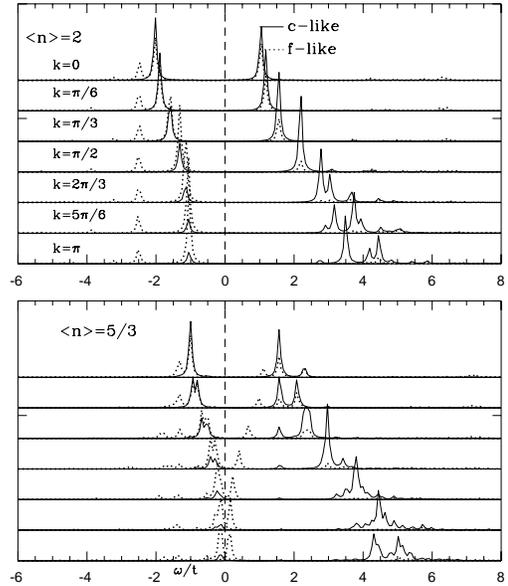}
\vspace{0cm}
\narrowtext
\caption{Lanczos spectrum for the  breathing $f$ orbital .
Parameter values are $U=\epsilon_f=2$, $V_1=2.0$,
$V_2=0.2$, $\epsilon_k=-2\cos(k)$.}
\label{fig14} 
\end{figure}
\noindent
charge gap: in our picture, the magnitude
of the gap is determined by the energy difference between
two cells with $2$ electrons on one hand, and
a cell with $1$ and one with
$3$ electrons on the other hand. Then, the electron
in the singly occupied cell can mix with the conduction band only
by using the (much smaller) hybridization
integral $V_2$, which corresponds to the  collapsed 
$f$ orbital. Only the single hole in the
cell occupied by $3$ electrons can delocalize
using the large hopping integral $V_1$.
A charge fluctuation thus will result in a huge loss
of {\em kinetic energy} and thus open a substantial
gap in the band structure
\begin{figure}[htb]
\epsfxsize=8.1cm
\vspace{-0.5cm}
\hspace{-0.0cm}\epsffile{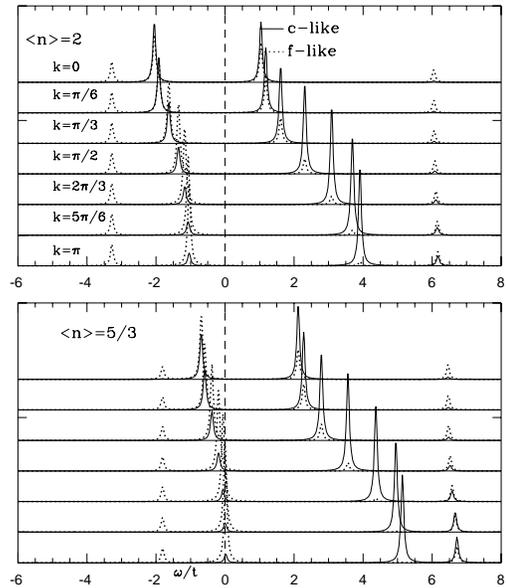}
\vspace{0cm}
\narrowtext
\caption{Theoretical spectra for the parameters of
Figure \ref{fig14}.}
\label{fig15} 
\end{figure}
\noindent
(the effect is thus similar,
but much stronger than the $U_{fc}$ discussed above).
This gap will be opened even in the complete absence of any Coulomb
repulsion on the $f$-orbital, so that the  breathing $f$-orbital  
can lead to
strong correlation-like behaviour even if there
is no really strong Coulomb repulsion at play.
In fact, comparing the Lanczos and theoretial results
(Figures \ref{fig14} and \ref{fig15})
one can note first of all a good agreement between the two,
and in addition a very substantial  correlation gap 
in the spectrum. Also, there are low-intensity `Hubbard bands',
whose position and spectral weight are reproduced
quite well by the theory. An interesting feature in the
Lanczos data is the extra
$f$-like band in the inverse
photoemission spectrum which disperses from
just above $E_F$ at $k$$=$$\pi$ to $\approx 2t$ above $E_F$
at $k$$=$$0$, and
which is completely absent in the theory. We believe that
this band is a consequence of 
superconducting pairing correlations in the
doped ground state. As discussed by various workers, the
breathing $f$-orbital presents a very strong
pairing mechanism\cite{Hirsch,Penc} for electrons in the
doped material, in that
the electrons will always pair up around one $f$-site
to hybridize by the large hopping parameter $V_1$. 
In the spectral function 
the presence of such pairing correlations, will manifest
itself as a Bogoliubov-type `mirror image' of the
topmost band, i.e. the replacement
\[
\tilde{\epsilon}_{\bbox{k}} - \mu \rightarrow
\pm \sqrt{ (\tilde{\epsilon}_{\bbox{k}}-\mu)^2 + \Delta_{sc}^2 }\;,
\]
where $\tilde{\epsilon}_{\bbox{k}}$ is the `nonsuperconducting dispersion' 
of the lower hybridization band (which should
correspond to the one given by our theory) and $\Delta_{sc}$
the superconducting gap. The latter appears to be quite
small in the numerical spectra. 
We defer a detailed discussion of superconductivity
to a separate publication,
but we note that our theory does not encompass
superconductivity, so that the absence of this band in the
theroretical spectra can be no surprise. \\
Summarizing this section, our theory
gives a remarkably good description of sometimes
rather complicated and unexpected features of 
a wide variety of extended versions of the
Kondo lattice. Moreover, it allows in all cases to extract simple
physical pictures in order to understand the overall trends.
The main inaccuracy concerned the positions of the
$f$-like Hubbard bands, but we believe that this is not
a severe deficiency because this is an extreme high-energy
feature. In the cases of extra intra-cell repulsions
the Lanczos spectra showed some indications of
a doping induced reconstruction of the electronic structure -
this could not be reproduced by our theory.
Also, superconductivity in the doped  `breathing $f$ model' 
naturally could not be described.
\section{Conclusion}
In summary, we have presented a `nearly analytical' theory for the Kondo 
lattice. In simplest terms it may be viewed as a Fermionic version 
of linear spin wave theory, in that we constructed
an effective Hamiltonian describing the pair creation and propagation
of Fermionic charge fluctuations on a strong coupling 
RVB-type vaccuum. This gives us a systematic way to
broaden the ionization and affinity states of a single cell
into bands, reminiscent of the cell perturbation theory
developed by Jefferson and co-workers\cite{Jefferson}.
For quite a number of different versions of the Kondo-lattice model,
the theoretical results for the single particle spectral function
were found to be in overall excellent agreement with exact diagonalization
of small clusters, as far as the dispersion of energy and spectral
weight is concerened. As a particularly encouraging
fact the rather nontrivial evolution of the
spectra with electron density in many cases is reproduced
in detail. This gives us some confidence that despite some
uncontrolled approximations the theory is already
 very close  to the correct picture. 
Given its extraordinary simplicity (all results are
obtained by an elementary 
Bogoliubov transformation) we believe that the theory easily
should be amenable to systematic improvement to describe as yet
neglected processes.\\
The emerging picture of the Kondo insulator then is
quite different from that of a conventional band insulator,
and in fact more reminiscent of a superconductor:
the ground state at half-filling corresponds to
an array of (overlapping) local singlets, where each Kondo-spin
forms a bound state with one conduction electron.
The system has a single particle gap for precisely the
same reason as in a superconductor: removing a conduction
electron from some site $i$ and reinserting it at a remote site $j$
breaks two $f$-$c$ pairs, whence there is an increase in energy of twice
the binding energy of a single pair.\\
The basic idea of the present work therefore is not restricted
to the Kondo lattice. Rather, once such an `RVB-vacuum'
has been identified, an essentially analogous construction can be
carried out also for other strongly correlated systems.
The most obvious example is the doped $t-J$ ladder, 
where the `rung singlet RVB state' may replace the
product of single cell states, and for which a
very similar construction can indeed been carried out\cite{spin}.\\
Perhaps the most important approximation we made is the neglect
of any excited states of two electrons in a cell. This means that
we have negelected single-cell states where one conduction electron
and one $f$-electron couple to a triplet. Such states
do in fact have a very small excitation energy $\propto V^2/U$.
As we have already mentioned, this restriction forbids the
propagating charge fluctuations to `radiate off'
triplet-like spin excitations
and thus makes their propagation completely coherent.
The good agreement with the numerics, and also the relatively
`coherent' nature of the Lanczos spectra themselves
(in which the incoherent
high-energy continua familiar from the $t-J$ or 
Hubbard model\cite{Dagoreview}
are almost completely absent), give us some confidence that despite
their low energy, the neglect of the singlet triplet excitations
is a good approximation as far as the single particle
properties are concerned. Moreover, it is quite easy to
incorporate these singlet-triplet excitations into
the present theory\cite{eos}: the three components of
a triplet state in cell $i$ can be grouped into
an SO(3) vector $\bbox{t}_i$. Then, we would model the
cell $i$ being in the $\alpha^{th}$ triplet state by the presence of
a {\em Bosonic} excitation, created by $t_{i,\alpha}^\dagger$.
In addition to a term describing the on-site energies
of these triplet we would obtain 
terms which describe the coupling between
the Fermionic charge fluctuations and the Bosonic spin
fluctuations. Their form can be inferred from rotational invariance:
for example 
$\bbox{t}_i^\dagger \cdot (a_{i,\tau}^\dagger \bbox{\sigma}_{\tau,\tau'} 
a_{j,\tau'}^{})$ (with $\bbox{\sigma}$ the vector of Pauli
matrices)
describes the hopping of a hole from $i\rightarrow j$ while
creating a spin excitation in cell $i$; the terms
$(\bbox{t}_j^\dagger \cdot \bbox{t}_i^{}) \cdot 
\sum_{\tau} a_{i,\tau}^\dagger a_{j,\tau}^{}$ and
$(a_{i,\tau}^\dagger \bbox{\sigma}_{\tau,\tau'} a_{j,\tau'}^{}) \cdot
(\bbox{t}_j^\dagger \times \bbox{t}_i^{})$ represent
two different ways for a spin excitation to exchange its position
with a hole. The actual prefactors of these terms can be computed
in an entirely analogous fashion as the hopping integrals for the
charge fluctuations themselves, and these terms can
be incorporated into the
formalism by using standard Green's function techniques.
Details will be reported elsewhere\cite{spin}.
Since already the simplest version of the theory apparently gives a
quite good description of the physics, one may expect
that such an extension is actually  `convergent'  and gives
a good description of the spin dynamics as well.\\
As a final remark, we note that the `vaccum' in the
case of the simple Kondo lattice was a unique
product of singlet states, and that the charge fluctuations
did have precisely the quantum numbers of electrons or holes.
For systems with a larger unit cell, however, one 
can envisage situations
where this is very different: one example would be a $2$-fold orbitally
degenerate $f$-level mixing with a nondegenerate
$c$-orbital.
For one electron/orbital the number of electrons/unit cell
then would be $3$, and the (Fermionic) charge fluctuations would
correspond to cells with an even number of electrons.
Thereby Hund's rule coupling between the degenerate $f$-orbitals
would favour high-spin states, so that one might actually
obtain `high-spin quasiparticles'. Form the good
success for the simple Kondo lattice one might expect that
the present formalism would continue to give
a good description even in this more complicated situation.\\
{\em Acknowledgement} 
This work was supported by Nederlands Stichting voor Fundamenteel
Onderzoek der Materie (FOM) and Stichting Scheikundig Onderzoek
Nederland (SON). Financial support of R. E. by the European
Community and of O. R. by the Soros Foundation is most gratefully
acknowledged.

\end{multicols}
\end{document}